\documentclass[reprint,twocolumn,showpacs,preprintnumbers,superscriptaddress,amsmath,amssymb]{revtex4-2}
\usepackage{dcolumn}
\usepackage{bm}
\usepackage[english]{babel}
\usepackage[strings]{underscore}
\usepackage{color} 
\usepackage{hyperref} 
\usepackage{graphicx}
\usepackage{siunitx}
\usepackage{xcolor}
\usepackage{amsmath,amssymb}
\usepackage[T1]{fontenc}
\usepackage{lmodern}
\usepackage[utf8]{inputenc}
\usepackage[english]{babel}
\usepackage{natbib}
\usepackage{hyperref}
\usepackage{mathtools}
\hypersetup{colorlinks=true,citecolor={blue},linkcolor={blue},urlcolor={blue}}
\usepackage{ulem}

\begin{document}
\title{Topology-Enabled Switchable Unidirectional Radiative Band in a Bilayer Photonic Crystal}

\author{Zhiyi Yuan}
\affiliation{
Centre for OptoElectronics and Biophotonics (COEB), School of Electrical and Electronic Engineering, Nanyang Technological University, Singapore 639798
}
\affiliation{CNRS-International-NTU-Thales Research Alliance (CINTRA), IRL 3288, Singapore 637553}
\affiliation{Institute of Materials Research and Engineering, Agency for Science Technology and Research (A*STAR), 2 Fusionopolis Way, Singapore 138634
}

\author{Vytautas Valuckas}
\affiliation{Institute of Materials Research and Engineering, Agency for Science Technology and Research (A*STAR), 2 Fusionopolis Way, Singapore 138634
}
\author{Yuhao Wang}
\affiliation{
Centre for OptoElectronics and Biophotonics (COEB), School of Electrical and Electronic Engineering, Nanyang Technological University, Singapore 639798
}
\affiliation{Institute of Materials Research and Engineering, Agency for Science Technology and Research (A*STAR), 2 Fusionopolis Way, Singapore 138634
}
\author{Thi Thu Ha Do}
\affiliation{Institute of Materials Research and Engineering, Agency for Science Technology and Research (A*STAR), 2 Fusionopolis Way, Singapore 138634
}
\author{Ningyuan Nie}
\affiliation{Centre for OptoElectronics and Biophotonics (COEB), School of Electrical and Electronic Engineering, Nanyang Technological University, Singapore 639798}
\author{Yu-Cheng Chen}
\affiliation{Centre for OptoElectronics and Biophotonics (COEB), School of Electrical and Electronic Engineering, Nanyang Technological University, Singapore 639798}
\author{Hai Son Nguyen}
\email{hai-son.nguyen@ec-lyon.fr}
\affiliation{CNRS-International-NTU-Thales Research Alliance (CINTRA), IRL 3288, Singapore 637553}
\affiliation{Ecole Centrale de Lyon, CNRS, INSA Lyon, \\ Universit\'e  Claude Bernard Lyon 1, CPE Lyon, CNRS, INL, UMR5270, Ecully 69130, France}
\author{Cuong Dang}
\email{hcdang@ntu.edu.sg}
\affiliation{
Centre for OptoElectronics and Biophotonics (COEB), School of Electrical and Electronic Engineering, Nanyang Technological University, Singapore 639798
}
\affiliation{CNRS-International-NTU-Thales Research Alliance (CINTRA), IRL 3288, Singapore 637553}
\author{Son Tung Ha}    
\email{ha_son_tung@imre.a-star.edu.sg}
\affiliation{Institute of Materials Research and Engineering, Agency for Science Technology and Research (A*STAR), 2 Fusionopolis Way, Singapore 138634
}


	\begin{abstract}
	Controlling how an open photonic system exchanges energy with its environment—and in particular how it radiates into the far field—is a cornerstone of non-Hermitian wave physics and a key enabler for directional photonic functionalities. Here, we propose a new route to robust unidirectional emission based on the non-Hermitian hybridization of resonances localized in spatially separated layers of a hetero-bilayer photonic crystal. By tailoring the interlayer coupling, we engineer hybrid photonc bands that exhibit strong unidirectional radiation across a broad spectral and momentum range while maintaining theoretically high quality factors. This asymmetric emission is organized by a topological vortex in a pseudo-polarization field defined from the front/back intensity imbalance, which endows the directionality with robustness against perturbations. We further show that, by tuning the surrounding refractive index, this singularity can be displaced in parameter space, enabling reversible switching of the emission direction and a reconfigurable far-field response. This framework opens perspectives for topological photonic sensing and for directional and switchable light sources, including unidirectional lasing supported by high-quality-factor modes.
	\end{abstract}
	
	\maketitle

    \emph{Introduction-} A central objective of nanophotonics is to shape how optical modes in open systems couple to the continuum, since this radiative coupling governs energy leakage, quality factors, and the angular distribution of emission \cite{Ha_2024,Kuznetsov_2024,Hsu_2016}. In periodic dielectric structures, this control is encoded in the dispersion and symmetry of guided resonances, which can be engineered through geometry but is ultimately constrained in single-layer platforms by limited degrees of freedom \cite{Hsu_2013, Hsu_2016,Qin_2025, Do_2025,Doiron_2024,Ha_2018,Yuan_2024}. Multilayer metasurfaces and photonic crystals (PhCs) overcome this restriction by introducing an additional vertical dimension for mode hybridization: stacking spatially separated resonant layers creates a coupled-mode space in which radiative channels can interfere, be redistributed between layers, and acquire new effective symmetries. This enlarged design space has enabled the realization of flatband dispersions\cite{Nguyen_2018,Huang_2022,Dong_2021,Nguyen_2022,Saadi_2024}, enhanced nonlinearities\cite{Zhou_2025,Qu_2025}, topological effects in synthetic momentum space\cite{Nguyen2023,Zhuang_2024,Canos_2025,Nguyen2026}, and polarization singularities\cite{Qin_2023,Kang_2025,Zhang_2023,Ni_2024}.

A particularly appealing consequence of this multilayer interference is the ability to engineer strong directional imbalance between the upward and downward radiation channels. This idea is captured by the recently introduced concept of unidirectional guided resonances (UGRs), which provides a route to mirror-free directional emission in photonic crystal slabs \cite{Yin_2020,Yin_2023,Zeng_2021}. UGRs arise when the coupling to the two half-spaces becomes strongly imbalanced, producing asymmetric far-field responses that can persist across a photonic band and has now been demonstrated in bilayer metasurfaces \cite{Zhuang_2024,Zhuang_2025,Choi_2025}. Despite this progress, existing UGR implementations face three recurring limitations: low quality factors due to intrinsically strong radiative leakage,  limited post-fabrication reversibility, and directionality predominantly at oblique angles rather than near-normal emission, which constrains their use in high-performance and reconfigurable devices.

    In this letter, we theoretically propose and experimentally demonstrate a new mechanism for achieving highly asymmetric radiation in a hetero-bilayer PhC, based on the non-Hermitian hybridization of symmetry-protected modes. Unlike previous UGRs, which rely on leaky resonances, the hybridized modes in our system originate from true bound states in the continuum (BICs) at the $\Gamma$ point, preserving their infinite Q-factors in the absence of symmetry breaking. Through careful engineering of the interlayer coupling, we demonstrate that the band corresponding to the bonding mode exhibits near-unity radiation asymmetry over a broad range of in-plane wavevectors, with light emitted predominantly in either the upward or downward direction. This unidirectionality is organized by a topological vortex in a synthetic parameter space, reflecting a phase singularity of a radiation-asymmetry pseudo-polarization field and arising from destructive interference that extinguishes radiation on one side. Crucially, the singularity is both robust and tunable: varying the surrounding refractive index (RI) displaces it continuously in parameter space, enabling reversible switching of the emission direction. Our results thus establish a general route to robust, reconfigurable, high-$Q$ directional emission, with perspectives for actively tunable metasurfaces and vertically emitting devices in integrated nanophotonic platforms.

	\begin{figure}
		\centering
		\includegraphics[width=0.8 \linewidth]{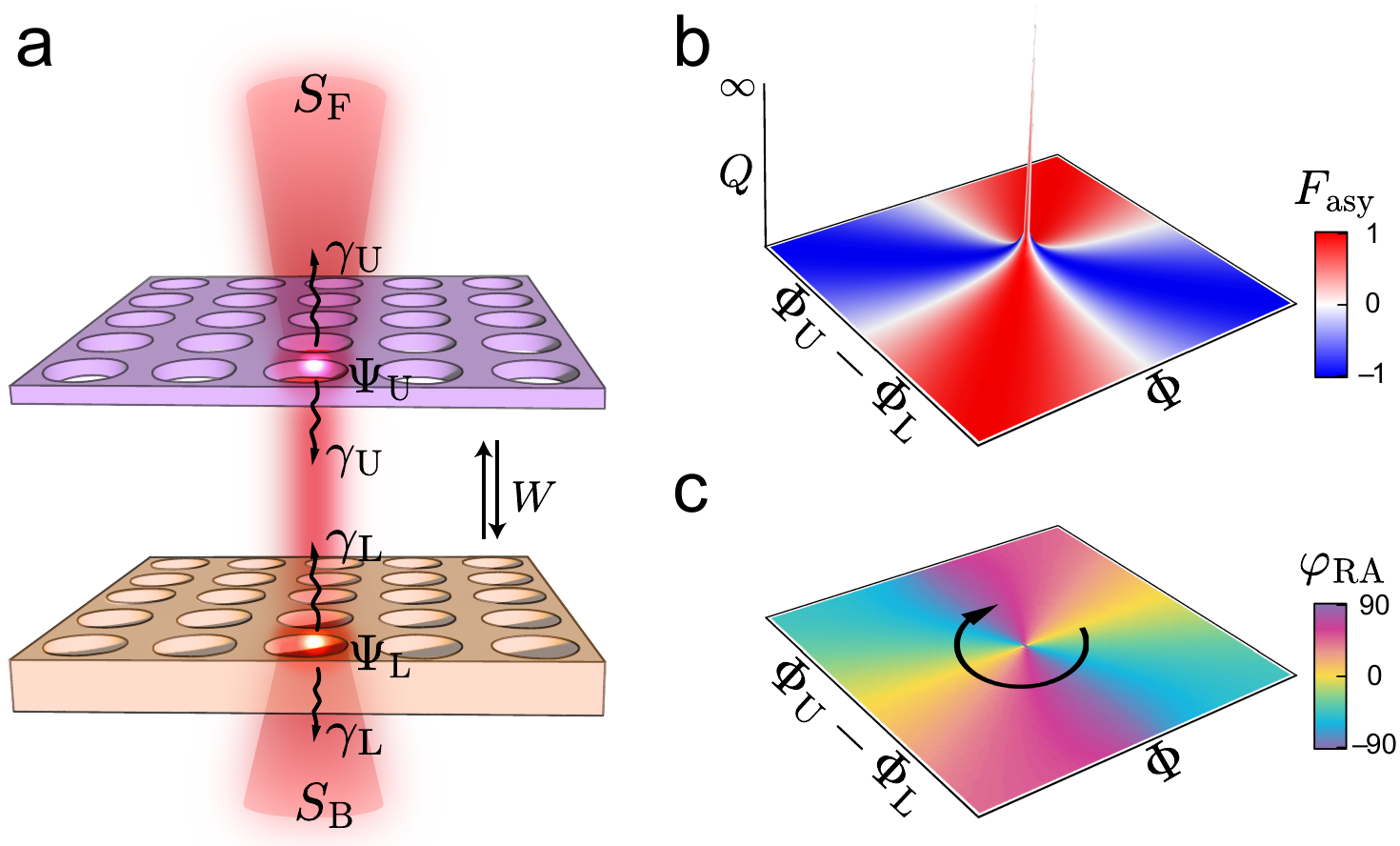}
		\caption{(a) Conceptual diagram of the hetero bilayer PhC with asymmetric emission. (b-c) Dependence of quality factor ($Q$) and asymmetry factor ($F_\mathrm{asy}$) (top panel) and pseudo-polarization angle (bottom panel) on the propagation phases in the bilayer PhC system. The results are analytically calculated from the simplified $2\times2$ TCMT model. $\Phi_{\mathrm{U}}$ and $\Phi_{\mathrm{L}}$ represent the phase factor caused by upper and lower PhC slabs, respectively. $\Phi$ is the phase factor induced by the interlayer gap. 
		}
		\label{Fig.1}
	\end{figure}
	
	\emph{Concept---}
The heterostructure consists of two vertically stacked PhC slabs made of materials with different refractive indices, separated by a subwavelength gap [see Fig.~\ref{Fig.1}(a)]. Each slab is vertically symmetric and supports a spectrally isolated guided resonance. In the basis of the uncoupled slab modes,  with modal amplitudes $\bm{\Psi}=\big(\Psi_\mathrm{U},\Psi_\mathrm{L}\big)^T$,
the hybridization is captured by the effective non-Hermitian Hamiltonian
\begin{equation}
H=
\begin{pmatrix}
\omega_\mathrm{U} & W \\
W & \omega_\mathrm{L}
\end{pmatrix}
+i
\begin{pmatrix}
\gamma_\mathrm{U} & e^{i(\Phi+\Phi_\mathrm{U})}\sqrt{\gamma_\mathrm{U}\gamma_\mathrm{L}} \\
e^{i(\Phi+\Phi_\mathrm{L})}\sqrt{\gamma_\mathrm{U}\gamma_\mathrm{L}} & \gamma_\mathrm{L}
\end{pmatrix},
\label{eq:H_two}
\end{equation}
which governs the temporal evolution $d\bm{\Psi}/dt=iH\bm{\Psi}$.
Here $\omega_m$ and $\gamma_m$ ($m=\mathrm{U,L}$) denote the resonance frequency and radiative decay rate of the isolated upper and lower slabs, respectively. While each slab radiates symmetrically by itself, stacking them introduces two distinct interlayer couplings:  a near-field (evanescent) coupling $W$ arising from modal overlap across the gap; and a radiative (continuum-mediated) coupling due to their shared access to the front and back radiation channels, represented by the off-diagonal imaginary terms $\propto \sqrt{\gamma_\mathrm{U}\gamma_\mathrm{L}}\,e^{i(\Phi+\Phi_{\mathrm{U,L}})}$ in Eq.~\eqref{eq:H_two}. The phase $\Phi$ is the propagation delay across the gap (set primarily by the gap thickness), whereas $\Phi_\mathrm{U}$ and $\Phi_\mathrm{L}$ are single-slab phase shifts determined by the respective slab thicknesses \cite{Alagappan_2024}. A full derivation of Eq.~\eqref{eq:H_two} from temporal coupled-mode theory is provided in the End Matter.

    The non-Hermitian Hamiltonian $H$ yields two eigenvectors $\bm{\Psi}_\pm=(\Psi_\mathrm{U}^\pm,\Psi_\mathrm{L}^\pm)^T$, with corresponding  complex eigenfrequency $\Omega_\pm=\omega_\pm+i\gamma_\pm$
    The coupling of these hybrid eigenmodes  to the upward and downward continua is generally unequal, enabling inherently asymmetric radiation.  Indeed, the far-field radiation amplitudes toward the front (superstrate) $S_\mathrm{F}$ and back (substrate) $S_\mathrm{B}$ are coherent superpositions of the fields emitted by the two slabs, given by $S_\mathrm{F} = \sqrt{\gamma_\mathrm{U}}\Psi_\mathrm{U}+\sqrt{\gamma_\mathrm{L}}\Psi_\mathrm{L}e^{i(\Phi+\Phi_\mathrm{U})},\quad S_\mathrm{B} = \sqrt{\gamma_\mathrm{L}}\Psi_\mathrm{L}+\sqrt{\gamma_\mathrm{U}}\Psi_\mathrm{U}e^{i(\Phi+\Phi_\mathrm{L})}$. The corresponding far-field intensities are $I_{\mathrm{F(B)}}=|S_{\mathrm{F(B)}}|^2$.

Complete suppression of radiation from an eigenmode ($I_\mathrm{F}=I_\mathrm{B}=0$) occurs when two conditions are simultaneously satisfied \cite{supp}: (i) the interlayer coupling $W$ fulfills the Friedrich--Wintgen (FW) interference condition $\frac{\omega_\mathrm{U}-\omega_\mathrm{L}}{W}=(-1)^n\left(\sqrt{\frac{\gamma_\mathrm{U}}{\gamma_\mathrm{L}}}-\sqrt{\frac{\gamma_\mathrm{L}}{\gamma_\mathrm{U}}}\right)$,
and (ii) the total phase delays obey $\Phi+\Phi_{\mathrm{U}}=\Phi+\Phi_{\mathrm{L}}=n\pi$, where $n$ is an integer\cite{supp}. Under these conditions, one hybrid eigenmode has a purely real eigenfrequency and becomes a Friedrich--Wintgen BIC (FW-BIC) \cite{Nguyen_2018,Letartre_2022,Li_2016}. Remarkably, even when the exact FW-BIC point is slightly detuned so that only one of the phase relations $\Phi+\Phi_{\mathrm{U}}=n\pi$ or $\Phi+\Phi_{\mathrm{L}}=n\pi$ holds, destructive interference can still cancel radiation on one side only, yielding highly asymmetric unidirectional emission: either $I_\mathrm{F}=0$ with $I_\mathrm{B}\neq 0$, or $I_\mathrm{B}=0$ with $I_\mathrm{F}\neq 0$. Thus, the three regimes: (1) total suppression (FW-BIC), (2) front-only emission, and (3) back-only emission, are unified manifestations of the same non-Hermitian hybridization and continuum-interference mechanism. Figure~\ref{Fig.1}(b) illustrates these regimes by mapping the $Q$ factor and the RA factor $F_{\mathrm{asy}}=(I_\mathrm{F}-I_\mathrm{B})/(I_\mathrm{F}+I_\mathrm{B})$ as a function of $\Phi$ and $\Phi_{\mathrm{U}}-\Phi_{\mathrm{L}}$ (while keeping $\Phi_{\mathrm{U}}=-\Phi_{\mathrm{L}}$), revealing the FW-BIC at $\Phi=\Phi_\mathrm{U}-\Phi_\mathrm{L}=0$ and unidirectional emission near the location of $\Phi+\Phi_{\mathrm{U,L}}=0$. We note that, in the special case of an identical bilayer ($\omega_\mathrm{U}=\omega_\mathrm{L}$ and $\gamma_\mathrm{U}=\gamma_\mathrm{L}$), the FW-BIC reduces to what is often termed a Fabry--P\'erot BIC: the Friedrich--Wintgen condition (i) is then automatically satisfied for any interlayer coupling $W$, and complete radiation cancellation is controlled solely by the total phase delay, i.e., condition (ii)  
\cite{Semushev_2025,Alagappan_2024,Letartre_2022}. In contrast, our hetero-bilayer design generally features finite detuning in both resonance frequencies and radiative decay rates. As a result, forming an FW-BIC requires simultaneous satisfaction of the phase condition (ii) and the coupling condition (i), the latter demanding precise engineering of the interlayer coupling strength $W$.

   From a topological perspective, the radiation asymmetry near the FW-BIC can be characterized by a pseudo-spin representation $(S_\mathrm{F},S_\mathrm{B})$, which forms the basis of the radiation-asymmetry (RA) pseudo-polarization framework \cite{supp,Zhuang_2024}. The associated pseudo-polarization vector is defined as:
\begin{equation}
\vec{\mathbf{S}} = \frac{1}{N}\Big[\vec{e}_x\!\left(S_\mathrm{F}+S_\mathrm{B}\right)+\vec{e}_y\!\left(S_\mathrm{F}-S_\mathrm{B}\right)\Big],\\
\label{eq:pseudo-polar-and-psi}
\end{equation}
   where $N=\sqrt{2\left(|S_\mathrm{F}|^2+|S_\mathrm{B}|^2\right)}$ is a normalization factor. Here the ``horizontal'' component $S_\mathrm{F}+S_\mathrm{B}$ corresponds to the symmetric radiation channel, whereas the ``vertical'' component $S_\mathrm{F}-S_\mathrm{B}$ corresponds to the antisymmetric channel. When an isolated FW-BIC occurs in a two-dimensional parameter space, both radiation components vanish ($S_\mathrm{F}=S_\mathrm{B}=0$). This makes the pseudo-polarization angle $\varphi_{\mathrm{RA}}$, given by $\tan\!\left(2\varphi_{\mathrm{RA}}\right) = \frac{|S_F|^2-|S_B|^2}{2\,\operatorname{Re}\!\left(S_F S_B^*\right)}$, undefined and becoming a vortex core. Upon encircling this singularity in parameter space, $\varphi_{\mathrm{RA}}$ winds by an integer multiple of $2\pi$, yielding the topological charge $q=\frac{1}{2\pi}\oint_{C}\nabla_{\mathbf{t}}\varphi_{\mathrm{RA}}(\mathbf{t})\cdot d\mathbf{t}$, with $\mathbf{t}=(\Phi_{\mathrm{U}}-\Phi_{\mathrm{L}},\,\Phi)$ [see Fig.~\ref{Fig.1}(c) with $q=-1$]. Consequently, the suppressed and unidirectional radiation states around the FW-BIC are not merely the result of fine-tuned interference, but are organized by a topological singularity rooted in non-Hermitian radiative coupling, which ensures robustness against perturbations. We stress that this RA singularity is not a genuine polarization singularity of the emitted field: the physical polarization remains essentially uniform over this parameter space, so the topological charge associated with the RA pseudo-polarization vortex in synthetic space can differ from that of the momentum-space polarization vortex, even though both originate from the same FW-BIC (see \cite{supp}).

\begin{figure}[h]
  \centering
  \includegraphics[width=0.48\textwidth]{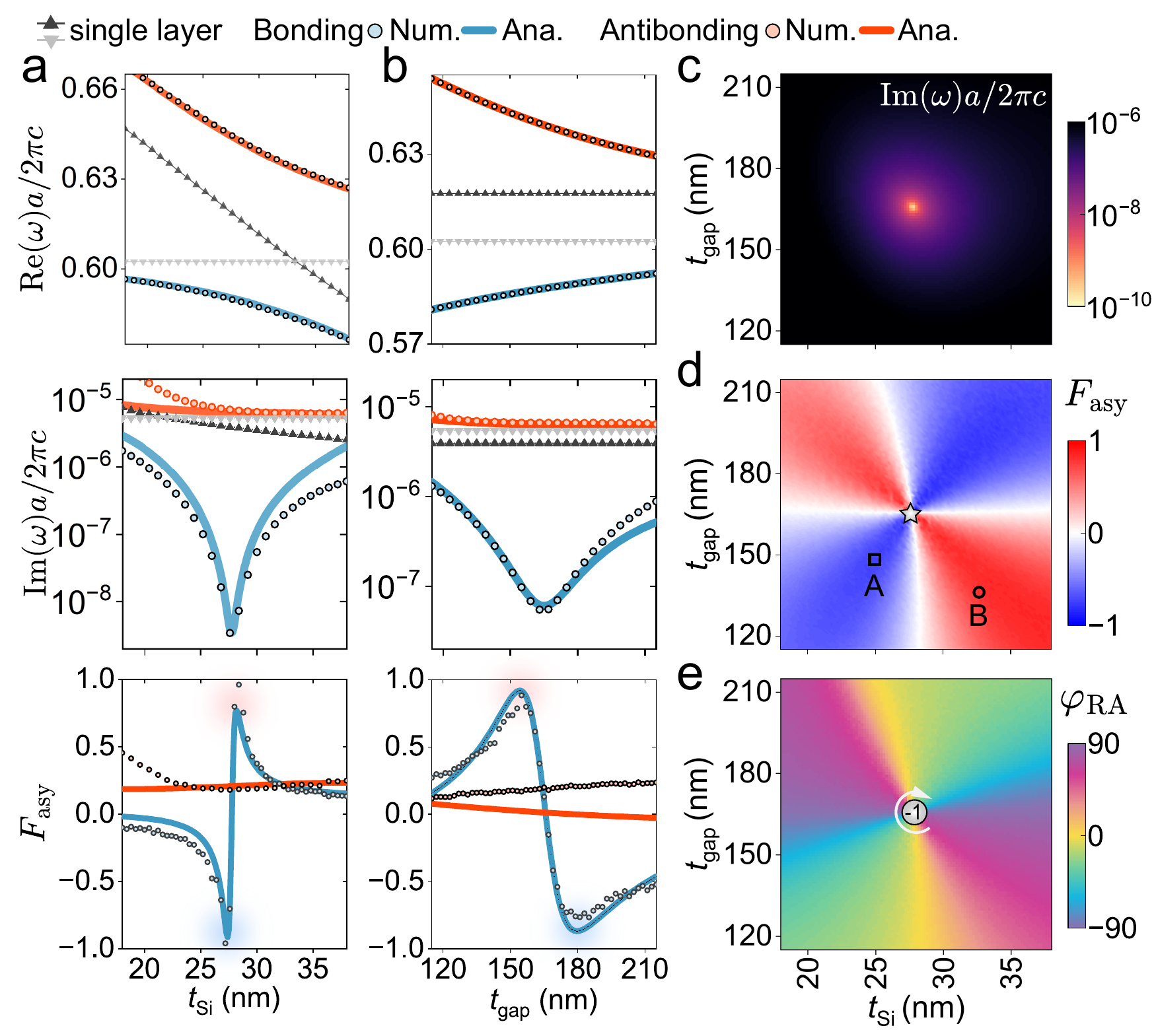}
  \caption{ (a-b) Bonding and antibonding mode behavior as a function of $t_{\mathrm{Si}}$ (a) and $t_{\mathrm{gap}}$ (b) for $k_x = 0.001\times2\pi/a$. Top: band energy. Middle: radiative decay rate. Bottom: RA factor. Solid lines denote the analytical calculation (i.e., ana.) results, and scattered dots denote the numerical simulation (i.e., num.) results. (c-e) Mapping of normalized radiative decay rate (c), RA factor $F_{\mathrm{asy}}$ (d), and pseudo-polarization angle $\varphi_{\mathrm{RA}}$ (e) for the bonding mode in the parameter space for $k_x = 0.001\times2\pi/a$. The square and circle labels in (d) represent parameters of the sample A and B, respectively in the experiments. In all simulations, the PhC is consist of a square lattice of air holes with a period  $a = $~\SI{445}{\nano\meter} and hole diameter  $d = $~\SI{300}{\nano\meter}.}
  \label{Fig.2}
\end{figure}

   \emph{Design and numerical results-}
   In a single-layer PhC slab with a square lattice of air holes, the photonic band structure typically comprises both leaky bands and BIC-hosting bands. In particular, symmetry-protected BICs occur at the $\Gamma$ point ($\mathbf{k}=0$) and can be classified by their near-field symmetry (e.g., magnetic-monopole-like and electric-quadrupole-like BICs), which dictates how they couple to the radiative continuum away from $\Gamma$ \cite{Ha_2018}. In a bilayer platform, non-Hermitian hybridization can in general occur either between two such BIC-derived resonances or between two leaky-band resonances; here we first focus on the BIC-derived case, which enables simultaneous directionality and high-$Q$ operation. As a case study, we consider a hetero-bilayer PhC composed of an amorphous silicon ($\mathrm{Si}$) upper slab, a titanium dioxide ($\mathrm{TiO_2}$) lower slab, and a silicon dioxide ($\mathrm{SiO_2}$) gap layer in between. Each PhC consists of a square lattice of air holes with identical lattice period and hole size, but differing slab thicknesses. The two slabs are designed such that their electric-quadrupole (EQ) BIC-derived bands are spectrally aligned.  When the two layers are brought into close proximity with a subwavelength gap, strong interlayer coupling occurs, leading to hybridization of the EQ-BIC modes into bonding and antibonding states (\cite{supp}). These hybrid modes exhibit antinode and node distributions along the vertical direction, respectively, while retaining the same in-plane symmetry of EQ-BICs.
   
The switching of unidirectional radiation enabled by non-Hermitian hybridization is examined through two parameter scans: (i) varying $t_{\mathrm{Si}}$ at fixed $t_{\mathrm{gap}}=\SI{163}{\nano\meter}$ and $t_{\mathrm{TiO_2}}=\SI{128}{\nano\meter}$, and (ii) varying $t_{\mathrm{gap}}$ at fixed $t_{\mathrm{Si}}=\SI{30}{\nano\meter}$ and $t_{\mathrm{TiO_2}}=\SI{128}{\nano\meter}$. Simulations are performed at a small oblique wavevector $k_x=0.001\times2\pi/a$ to weakly lift the EQ-BIC symmetry protection and allow controlled radiation. Tuning $t_{\mathrm{Si}}$ modifies both the detuning $\omega_\mathrm{U}-\omega_\mathrm{L}$ and the phase mismatch $\Phi_\mathrm{U}-\Phi_\mathrm{L}$, strongly impacting the hybrid-mode energies, losses, and RA factor. As shown in Fig.\ref{Fig.2}(a) (top), an avoided crossing evidences strong coupling. The radiative losses are exchanged between the modes [see Fig.\ref{Fig.2}(a), middle], and the bonding mode becomes nearly lossless at $t_{\mathrm{Si}}=\SI{27.6}{\nano\meter}$, marking an FW-BIC and a $Q$ enhancement of up to three orders of magnitude. Crossing this point flips $F_{\mathrm{asy}}$ from $\approx-1$ to $\approx+1$ [see Fig.\ref{Fig.2}(a), bottom], i.e., the emission switches from back to front, because near $\Phi_\mathrm{U}\approx\Phi_\mathrm{L}$ the conditions $\Phi+\Phi_\mathrm{U}=n\pi$ and $\Phi+\Phi_\mathrm{L}=n\pi$ are satisfied on opposite sides of the FW-BIC, enabling single-sided cancellation alternately in the two directions. A similar behavior occurs when varying $t_{\mathrm{gap}}$ [see Fig.\ref{Fig.2}(b)], which tunes $W$ and $\Phi$: increasing the gap reduces the bonding–antibonding splitting [see Fig.~\ref{Fig.2}(b), top] and produces an abrupt change in losses and $F_{\mathrm{asy}}$ near the critical thickness. All numerical observations in the two parameter scans are quantitatively captured by our coupled-mode analytical model \cite{supp}. Although the scans are shown at a single small oblique wavevector $k_x=0.001\times2\pi/a$, the RA factor $F_{\mathrm{asy}}$ remains near-unity over a broad range of in-plane momenta, yielding unidirectional bands rather than isolated points (see next section). This differs from most prior works, where unidirectionality is restricted to a single momentum \cite{Yin_2020,Yin_2023,Zhou_2016} or to one propagation direction only \cite{Choi_2025}.

To highlight  the topological organization of the radiation asymmetry, we map the radiative decay rate, $F_{\mathrm{asy}}$, and the RA phase $\varphi_{\mathrm{RA}}$ of the bonding mode over parameter space [see Figs.\ref{Fig.2}(c–e)]. The decay rate vanishes at an isolated point, confirming an FW-BIC [see Fig.\ref{Fig.2}(c)]. Around it, $F_{\mathrm{asy}}$ forms a characteristic four-quadrant pattern [see Fig.\ref{Fig.2}(d)] arising from $|S_\mathrm{F}|^2\neq|S_\mathrm{B}|^2$. The corresponding $\varphi_{\mathrm{RA}}$ map winds around the FW-BIC [see Fig.\ref{Fig.2}(e)], revealing a radiation-asymmetry singularity with topological charge $q=-1$ pinned to the FW-BIC.

   Furthermore, while our discussion so far has focused on high-$Q$ modes such as the EQ-BIC band, the same mechanism for the emergence of asymmetric radiation also applies to low-$Q$ leaky modes in the bilayer PhC\cite{supp}, thereby extending the applicability of the concept to regimes requiring strong radiation-coupling efficiency.\\

	\begin{figure}[h] 
		\centering
		\includegraphics[width=0.48 \textwidth]{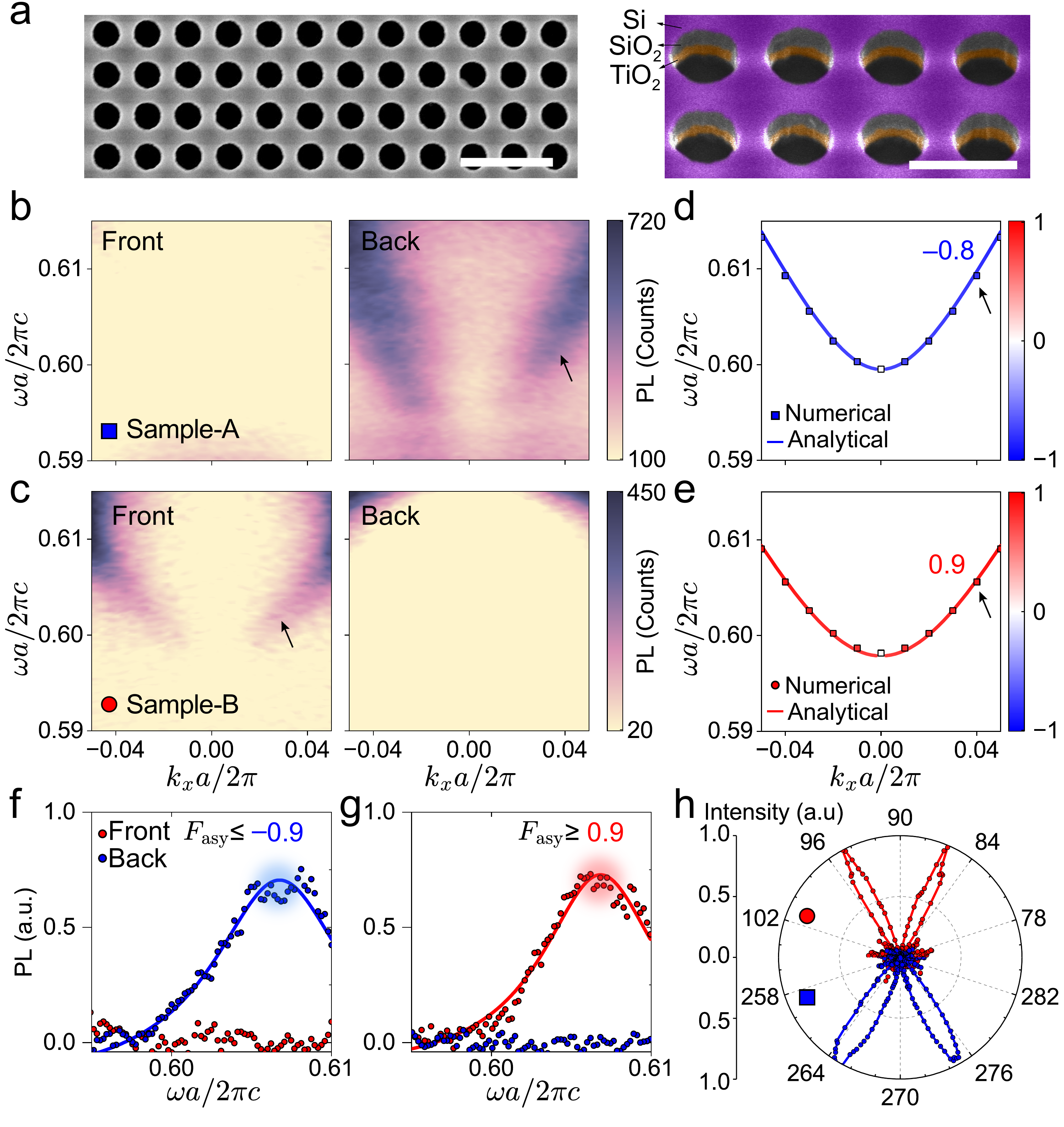}
		\caption{(a) SEM images (left: top-view; right: tilted view) of the fabricated bilayer PhC. False colors represent different layers: purple –$\mathrm{Si}$; gray –$\mathrm{SiO_2}$; yellow –$\mathrm{TiO_2}$. The scale bar is \SI{500}{\nano\meter}. (b-c) Measured momentum-resolved PL spectra of the sample A (b) and sample B (c). Left: front-side. Right: back-side. (d-e) Numerical and analytical simulated band RA of the EQ-BIC bonding mode in sample-A (d) and sample-B (e). The black arrow indicates the RA factor at $k_x = 0.04\times2\/a$. (f-g) Comparison of the PL spectra collected from front and back sides at $k_x = 0.04\times2\/a$, from sample A (f) and sample B (g). (h) Emission directionality of EQ-BIC bonding mode at normalized angular frequency $\sim0.605$ from sample-A (blue) and sample-B (red), after subtracting the contributions from larger-angle leaky mode emissions. Specifically, sample A features $t_{\mathrm{gap}}=\SI{144}{\nano\meter}$ and $t_{\mathrm{Si}}=\SI{24}{\nano\meter}$, while sample B has $t_{\mathrm{gap}}=\SI{130}{\nano\meter}$ and $t_{\mathrm{Si}}=\SI{32}{\nano\meter}$. } 
		\label{Fig.3}
	\end{figure}    
    
	\emph{Experimental demonstration of unidirectional radiation band in bilayer PhC-} To verify our theoretical and numerical findings, we fabricate the hetero-bilayer PhC (Fig. \ref{Fig.3}(a)(see \cite{supp} for details). Subsequently, the samples are coated with a thin PMMA dye layer, which has a RI similar to that of the quartz substrate and the $\mathrm{SiO_2}$ gap layer ($\sim 1.47$). Two samples, named A and B, are designed to exhibit unidirectional emission toward the back and front sides, respectively, as discussed in the previous section and indicated in Fig.~\ref{Fig.2}(d). The optical properties of the samples are characterized using a custom-built back-focal-plane (BFP) micro-spectrometer setup \cite{supp}, enabling angle-resolved photoluminescence (PL) and reflectivity ($R$) measurements.

    To evaluate the radiation asymmetry of the bonding mode, we measured angle-resolved PL spectra from both the front and back sides of each sample (Figs.~\ref{Fig.3}(b) and \ref{Fig.3}(c); a larger spectral range is shown in the SM). For sample~A, front-side emission is negligible, with nearly all PL emitted from the back side [see Fig.~\ref{Fig.3}(b)]. Conversely, sample B exhibits negligible back-side emission, with PL predominantly emitted from the front side [see Fig.~\ref{Fig.3}(c)]. In both cases, no emission is observed at the $\Gamma$ point, consistent with the symmetry-protected BIC nature of the bonding mode. This pronounced but opposite directional asymmetry in emission from the two samples agrees well with both our numerical simulations and analytical predictions [see Figs.~\ref{Fig.3}(d) and \ref{Fig.3}(e)]. Angle-resolved $R$ spectroscopy further confirms these observations~\cite{supp}. Moreover, analytical calculations based on our full-mode theoretical model show excellent agreement with FEM simulations~\cite{supp}. Quantitative evaluation of the radiation asymmetry was performed by extracting the PL spectra at a fixed in-plane momentum. As an example, Figs.~3f and 3g present the front- and back-side PL spectra at $k_x=0.04\times 2\pi/a$ for samples~A and B. These measurements reveal nearly perfect unidirectional emission, with the RA factor $F_{\mathrm{asy}}$ approaching $\pm1$. The unidirectional nature of the emission is further confirmed by analyzing the far-field radiation patterns at a fixed frequency. As shown in Fig.~\ref{Fig.3}(h), the $xz$ plane radiation patterns at a normalized angular frequency of approximately 0.605 display almost perfect unidirectionality for both samples, consistent with the momentum-resolved PL measurements. These results unambiguously confirm that the engineered interlayer coupling enables robust and highly directional emission, in excellent agreement with our theoretical predictions.

	\begin{figure}[h]
		\centering
		\includegraphics[width=0.48\textwidth]{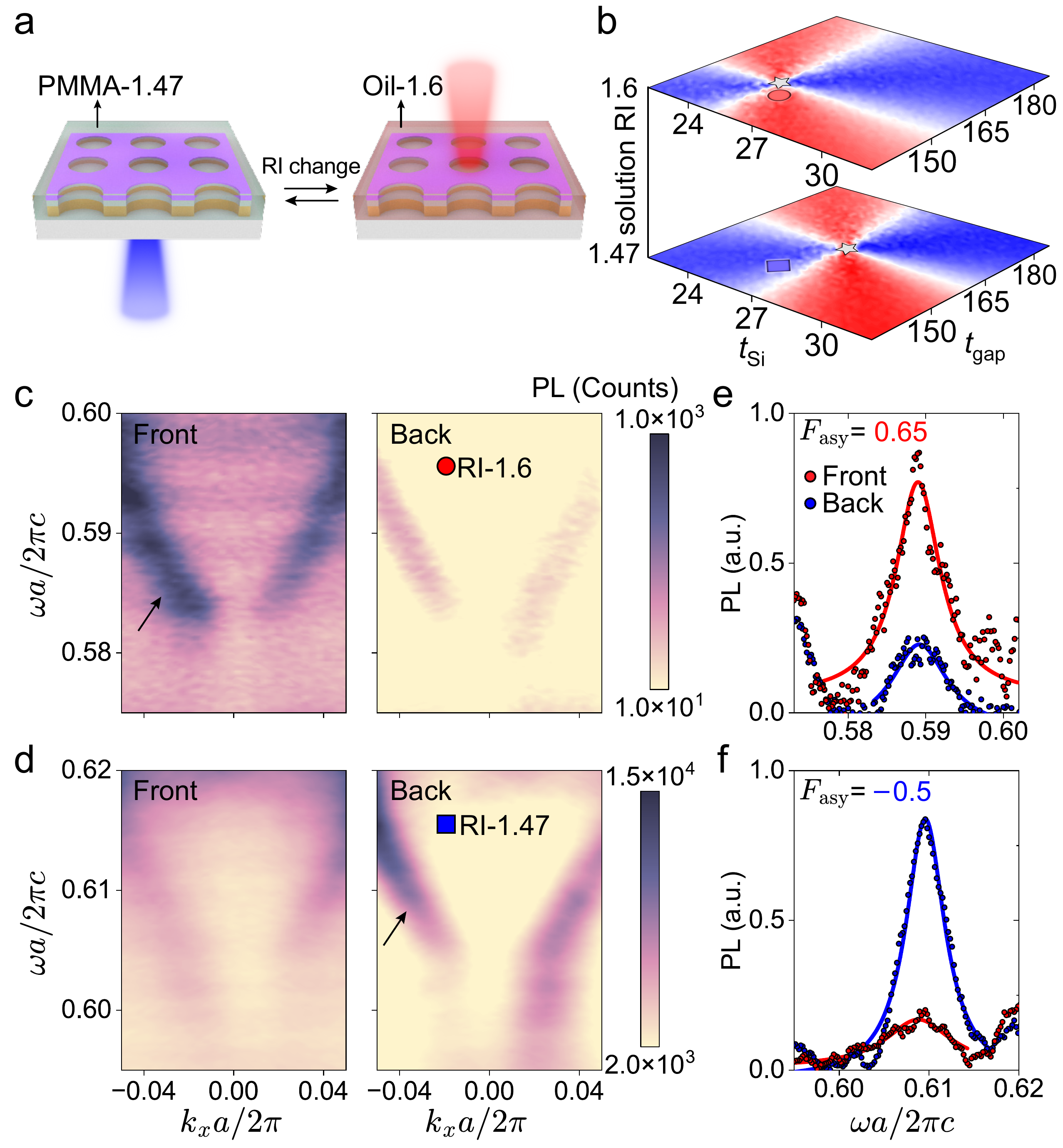} 
		\caption{(a) Schematic diagram of a tunable radiation-asymmetry device. By changing the surrounding solution, the refractive indices of the superstrate and nanoholes are adjusted. (b) Mapping of the RA factor in the parameter space under varying RI. The singularity shifts within the parameter space as the the solution RI changes. The red (blue) circle indicates the tunable sample in an oil (PMMA) environment. (c,d) Measured angle-resolved PL spectra of the tunable sample in the oil (c) and in PMMA (d). Left: front side. Right: back side. (e,f) Comparison of front-side and back-side PL spectra at a specific momentum ($k_x=0.03\times 2\pi/a$), measured in oil (e) and in PMMA (f). Sample C features $t_{\mathrm{gap}}=\SI{144}{\nano\meter}$ and $t_{\mathrm{Si}}=\SI{26}{\nano\meter}$}
		\label{Fig.4} 
	\end{figure}
	
	\emph{Switchable highly asymmetric radiation -} 
    We now investigate the tunability of the radiation-asymmetric BIC mode by modifying the surrounding RI. To demonstrate switchable emission directionality, we immerse the PhC sample in a liquid medium containing the same dye but with a different RI [see Fig.~\ref{Fig.4}(a)]. In this configuration, the RI of the substrate remains unchanged, while the RIs of the superstrate and nanoholes are altered. Owing to the topological robustness of the radiation asymmetry singularity, the four-quadrant pattern of the asymmetry factor $F_{\mathrm{asy}}$ persists under environmental RI changes, as confirmed by numerical simulations when the immersion medium is changed from PMMA ($n=1.47$) to oil ($n=1.6$), corresponding to a relatively small RI variation $\Delta n=0.13$ [see Fig.~\ref{Fig.4}(b)]. Although the singularity is displaced within the parameter space, its topological nature remains intact, resulting in a rigid translation of the $F_{\mathrm{asy}}$ patterns. Based on these insights, we design a tunable sample C, this design enables the emission direction to switch from negative to positive radiation asymmetry as the surrounding RI changes. Specifically, the operating point moves from the region marked by a blue square (negative asymmetry) to the region marked by a red circle (positive asymmetry) in the $F_{\mathrm{asy}}$ maps. These provide design rules for a tunable sample C whose design is labeled in Fig.~\ref{Fig.4}(b). Such design allows the emission of sample C located switch from negative radiation asymmetry [marked as the blue square in Fig.~\ref{Fig.4}(b)] to positive asymmetry [marked as the red circle in in Fig.~\ref{Fig.4}(b)]. Experimental angle-resolved PL spectra of sample~C in two different environments are shown in Figs.~\ref{Fig.4}(c) and \ref{Fig.4}(d). These measurements clearly demonstrate a reversal of the radiation asymmetry of the EQ-BIC bonding band. This switching behavior is accompanied by a slight redshift of the emission band, in excellent agreement with numerical simulations\cite{supp}. To better illustrate the change in directionality, we extract the PL spectra at a representative in-plane momentum $k_x=0.03\times 2\pi/a$, as shown in Figs.~\ref{Fig.4}(e) and \ref{Fig.4}(f) after baseline subtraction. By integrating the peak areas, we obtain RA factors of $F_{\mathrm{asy}}=0.65$ and $F_{\mathrm{asy}}=-0.5$, confirming the reversal of the emission direction. Theoretical predictions from both our analytical model and FEM simulations~\cite{supp}, yield corresponding values of $F_{\mathrm{asy}}=0.66$ and $F_{\mathrm{asy}}=-0.47$, in excellent agreement with the experimental data.


\emph{Conclusion-}
We have demonstrated that robust directionality can emerge as a consequence of the topology of radiative coupling in an open system, rather than from fine-tuned geometric asymmetry. In a hetero-bilayer PhC, non-Hermitian hybridization of guided resonances localized in different layers redistributes coupling to the upward and downward continua, producing near-unity radiation asymmetry over an extended spectral and momentum range while retaining high-$Q$ behavior associated with symmetry-protected BICs. The resulting emission landscape is governed by a radiation-asymmetry singularity---a vortex of the pseudo-polarization angle defined by the front/back imbalance---which anchors the transitions between suppressed, balanced, and unidirectional radiation. Because this singularity carries a quantized topological charge, the directional response is resilient to perturbations and admits a systematic design strategy based on moving the singularity in a synthetic parameter space. We further show that environmental refractive-index tuning provides a reversible handle to translate the singularity and thereby switch the emission direction. These findings position multilayer PhCs as a minimal and scalable platform to engineer topology in the far field of non-Hermitian photonics, opening avenues for reconfigurable directional emitters, topological sensors that rely on intensity imbalance rather than spectral readout, and unidirectional lasing supported by high-$Q$ modes.

	\begin{acknowledgments}
    S.T.H, T.T.H.D and V.V acknowledge funding support from A*STAR MTC-Programmatic Fund (M24N9b0122) and National Research Foundation, Singapore – Frontier Competitive Research Grant (NRF-F-CRP-2024-0009). Z. Y., C. D., acknowledge funding support from the National Research Foundation Competitive Research Program (NRF-CRP29-2022-0003), Ministry of Education, Singapore under its AcRF Tier 2 grant (MOE-T2EP50121-0012), and AcRF Tier 1 grant (RG140/23). This project is partly funded by the French  National Research Agency (ANR), STRONG-NANO (ANR-22-CE24-0025), POLAROID (ANR-24-CE24-7616-01), SUPER-HERO (ANR-25-CE24-4066), and CHOCOLAT (ANR- 44225-CE47-6432)
	\end{acknowledgments}

	\bibliography{Citations}

    \onecolumngrid
    \section*{End Matter}
    \twocolumngrid

  \subsection{Temporal coupled model theory for bilayer photonic crystals}

\begin{figure}[htb]
\centering
\includegraphics[width=0.4\linewidth]{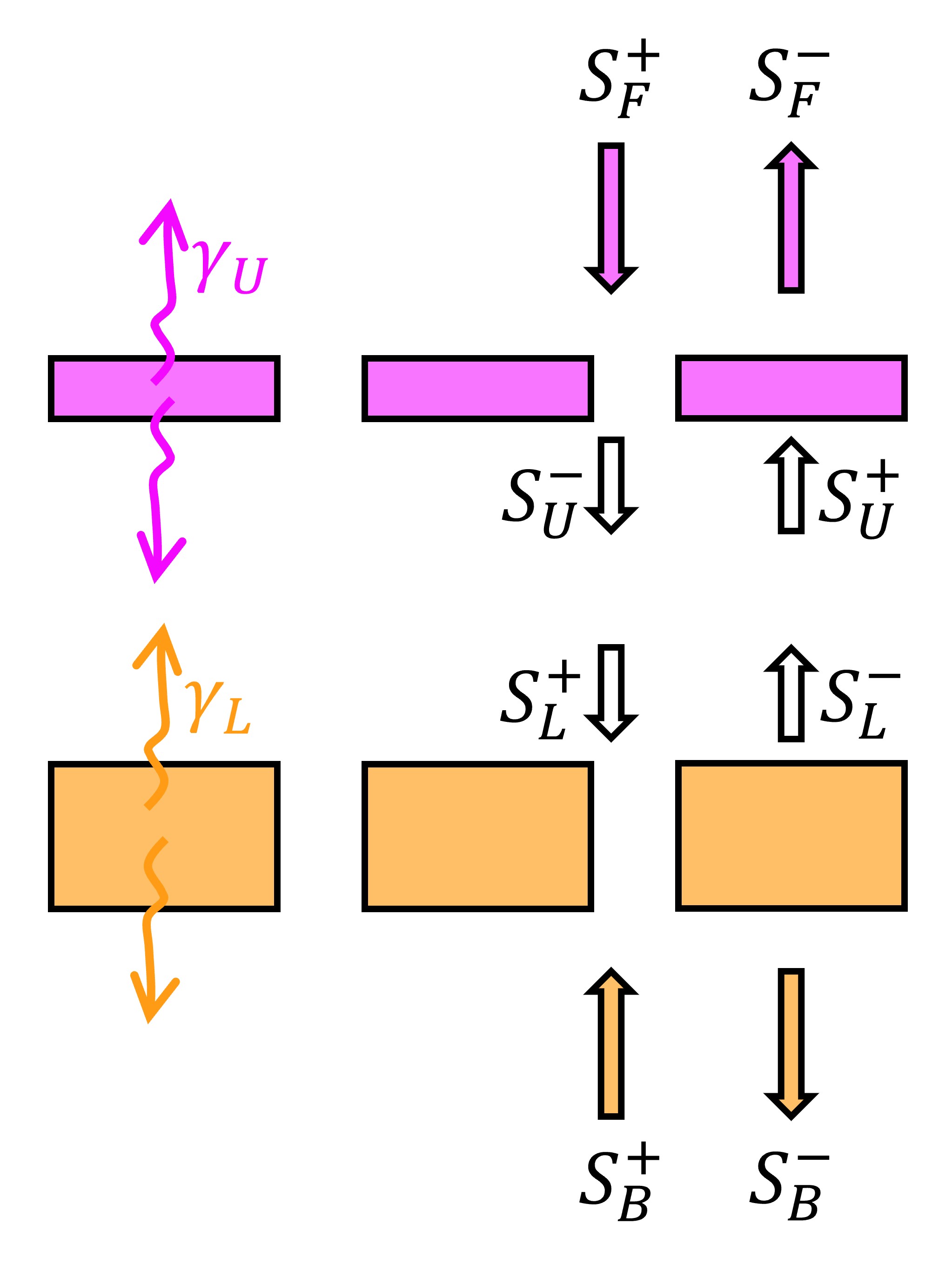}
\caption{Schematic of the bilayer PhC slabs and the four radiation ports used in the coupled-mode description.}
\label{Fig.5}
\end{figure}

We model the bilayer as two vertically symmetric PhC slabs separated by a gap of thickness $t_{\mathrm{gap}}$ [see Fig.~\ref{Fig.5}]. The structure is coupled to four radiation ports: $S_\mathrm{F}$ and $S_\mathrm{B}$ denote the fields in the superstrate and substrate continua, respectively, while $S_\mathrm{U}$ and $S_\mathrm{L}$ represent the forward/backward propagating fields inside the spacer, below the upper slab and above the lower slab. In the spectral range of interest, the upper (lower) slab supports a single Bloch resonance with frequency $\omega_\mathrm{U}$ ($\omega_\mathrm{L}$) and radiative decay rate $\gamma_\mathrm{U}$ ($\gamma_\mathrm{L}$). We assume that both resonances are even under vertical mirror symmetry (TE-like), so that each slab couples symmetrically to the upward and downward channels.

Let $\Psi_\mathrm{U}$ and $\Psi_\mathrm{L}$ be the modal amplitudes of the resonances in the upper and lower slabs. Temporal coupled-mode theory (TCMT) relates these amplitudes to the incoming/outgoing fields at the ports \cite{Fan_2003,Suh_2004,Zhou_2016}:
\begin{equation}
\frac{d\Psi_\mathrm{U}}{dt}
= i(\omega_\mathrm{U}+i\gamma_\mathrm{U})\Psi_\mathrm{U}
+ iW\Psi_\mathrm{L}
+ k_\mathrm{U} S_\mathrm{F}^+ + k_\mathrm{U} S_\mathrm{U}^+ ,
\label{eq:a1}
\end{equation}
\begin{equation}
\frac{d\Psi_\mathrm{L}}{dt}
= i(\omega_\mathrm{L}+i\gamma_\mathrm{L})\Psi_\mathrm{L}
+ iW\Psi_\mathrm{U}
+ k_\mathrm{L} S_\mathrm{B}^+ + k_\mathrm{L} S_\mathrm{L}^+ ,
\label{eq:a2}
\end{equation}
where $W$ is the near-field (evanescent) coupling between the two resonances, and $k_m$ ($m=\mathrm{U,L}$) are coupling coefficients from the input channels to the resonances.

The outgoing fields are related to the incoming fields through direct (non-resonant) transmission across each slab and resonant re-radiation:
\begin{equation}
\begin{pmatrix}
S_\mathrm{F}^-\\
S_\mathrm{U}^-
\end{pmatrix}
=
\begin{pmatrix}
0 & e^{i\Phi_\mathrm{U}}\\
e^{i\Phi_\mathrm{U}} & 0
\end{pmatrix}
\begin{pmatrix}
S_\mathrm{F}^+\\
S_\mathrm{U}^+
\end{pmatrix}
+
d_\mathrm{U}
\begin{pmatrix}
\Psi_\mathrm{U}\\
\Psi_\mathrm{U}
\end{pmatrix},
\label{eq:s1-s2}
\end{equation}
\begin{equation}
\begin{pmatrix}
S_\mathrm{B}^-\\
S_\mathrm{L}^-
\end{pmatrix}
=
\begin{pmatrix}
0 & e^{i\Phi_\mathrm{L}}\\
e^{i\Phi_\mathrm{L}} & 0
\end{pmatrix}
\begin{pmatrix}
S_\mathrm{B}^+\\
S_\mathrm{L}^+
\end{pmatrix}
+
d_\mathrm{L}
\begin{pmatrix}
\Psi_\mathrm{L}\\
\Psi_\mathrm{L}
\end{pmatrix},
\label{eq:s3-s4}
\end{equation}
where $\Phi_\mathrm{U}$ and $\Phi_\mathrm{L}$ are the single-slab propagation phases, and $d_m$ are the resonance-to-port coupling coefficients. As required by energy conservation and time-reversal symmetry, $k_m$ and $d_m$ are not independent \cite{Alagappan_2024}.

Propagation through the spacer couples the internal ports $S_\mathrm{U}$ and $S_\mathrm{L}$. For a uniform gap, the forward and backward waves acquire a phase $\Phi$ set by $t_{\mathrm{gap}}$:
\begin{equation}
S_\mathrm{L}^+ = S_\mathrm{U}^- e^{i\Phi},
\qquad
S_\mathrm{U}^+ = S_\mathrm{L}^- e^{i\Phi}.
\label{eq:propagate}
\end{equation}

We now focus on radiation from the coupled resonances, i.e., no external illumination from the superstrate or substrate ($S_\mathrm{F}^+=S_\mathrm{B}^+=0$). Eliminating the internal ports using Eqs.~\eqref{eq:s1-s2}--\eqref{eq:propagate} yields a closed evolution equation for $\bm{\Psi}=(\Psi_\mathrm{U},\Psi_\mathrm{L})^T$:
\begin{equation}
\frac{d}{dt}
\begin{pmatrix}
\Psi_\mathrm{U}\\
\Psi_\mathrm{L}
\end{pmatrix}
= iH
\begin{pmatrix}
\Psi_\mathrm{U}\\
\Psi_\mathrm{L}
\end{pmatrix},
\end{equation}
with the effective non-Hermitian Hamiltonian
\begin{equation}
H=
\begin{pmatrix}
\omega_\mathrm{U} & W \\
W & \omega_\mathrm{L}
\end{pmatrix}
+i
\begin{pmatrix}
\gamma_\mathrm{U} & e^{i(\Phi+\Phi_\mathrm{U})}\sqrt{\gamma_\mathrm{U}\gamma_\mathrm{L}} \\
e^{i(\Phi+\Phi_\mathrm{L})}\sqrt{\gamma_\mathrm{U}\gamma_\mathrm{L}} & \gamma_\mathrm{L}
\end{pmatrix}.
\label{eq:H_twotwo}
\end{equation}

The hybrid modes are obtained by diagonalizing $H$, which gives complex eigenfrequencies $\Omega_\pm=\omega_\pm+i\gamma_\pm$ and eigenvectors $\bm{\Psi}_\pm=(\Psi_\mathrm{U}^\pm,\Psi_\mathrm{L}^\pm)^T$. Their far-field radiation into the superstrate and substrate is given by the outgoing fluxes $S_\mathrm{F}^-$ and $S_\mathrm{B}^-$. Using $S_\mathrm{F}^+=S_\mathrm{B}^+=0$ in Eqs.~\eqref{eq:s1-s2}--\eqref{eq:propagate}, we obtain
\begin{equation}
S_{\mathrm{F}\pm}^- =
\sqrt{\gamma_\mathrm{U}}\Psi_\mathrm{U}^\pm
+
\sqrt{\gamma_\mathrm{L}}\Psi_\mathrm{L}^\pm e^{-i(\Phi+\Phi_\mathrm{U})},
\end{equation}
\begin{equation}
S_{\mathrm{B}\pm}^- =
\sqrt{\gamma_\mathrm{L}}\Psi_\mathrm{L}^\pm
+
\sqrt{\gamma_\mathrm{U}}\Psi_\mathrm{U}^\pm e^{-i(\Phi+\Phi_\mathrm{L})}.
\end{equation}
The corresponding intensities are
\begin{equation}
I_{\mathrm{F}\pm}=|S_{\mathrm{F}\pm}^-|^2,
\qquad
I_{\mathrm{B}\pm}=|S_{\mathrm{B}\pm}^-|^2,
\label{eq:farfield_+-}
\end{equation}
with total radiation $I_\pm=I_{\mathrm{F}\pm}+I_{\mathrm{B}\pm}$ and radiation-asymmetry factor
\begin{equation}
F_{\mathrm{asy}}^\pm
=
\frac{I_{\mathrm{F}\pm}-I_{\mathrm{B}\pm}}{I_{\mathrm{F}\pm}+I_{\mathrm{B}\pm}}.
\label{eq:asy}
\end{equation}
These expressions are used to compute the normalized loss rate and the RA factor reported in the main text.

\subsection{Experimental $Q$ factors}
\begin{figure}[htb!]
\centering
\includegraphics[width=1\linewidth]{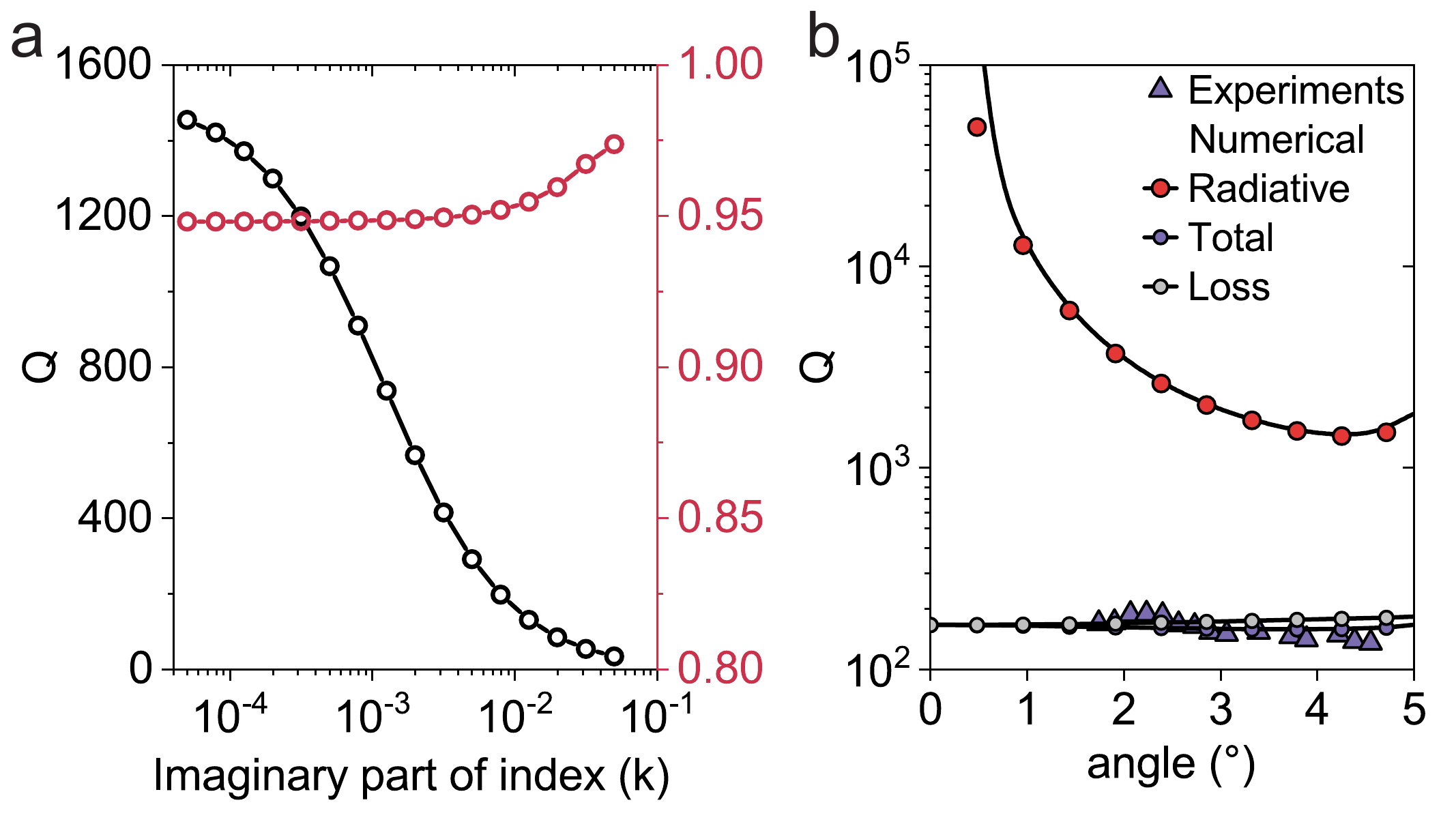}
\caption{(a) Total $Q$ factor and RA factor versus the imaginary part of the environmental refractive index, evaluated at $k_x=0.05\times2\pi/a$. (b) Experimental and numerical $Q$ factors versus in-plane momentum (angle). The experimental $Q$ is limited by nonradiative loss.}
\label{Fig.6}
\end{figure}

The measured linewidth is set by both radiative and nonradiative channels, such that
$1/Q_{\mathrm{total}}=1/Q_{\mathrm{rad}}+1/Q_{\mathrm{nonrad}}$.
As the imaginary part of the environmental refractive index increases, $Q_{\mathrm{total}}$ drops rapidly [see Fig.~\ref{Fig.6}(a)], whereas the RA factor remains nearly unchanged over a broad range of absorption. Although the targeted BIC-derived mode has an ideal diverging $Q$ at the $\Gamma$ point (and a decreasing $Q$ away from $\Gamma$), the experimental values remain low ($\sim 200$) at small angles because $Q_{\mathrm{nonrad}}$ is dominated by material absorption (PMMA, dye, and $\mathrm{Si}$) and fabrication imperfections.

\subsection{Future experimental schemes}

While the PMMA environment ($n=1.47$) yields moderate asymmetry, stronger unidirectionality is expected at lower refractive indices; for example, simulations give $F_{\mathrm{asy}}=-0.75$ at $n=1.30$ \cite{supp}. The sensing response can be further amplified by selectively filling the nanoholes and/or the spacer with a higher-index material (e.g., $\mathrm{SiO_2}$) \cite{supp}. In this configuration, even a small RI change from $n=1.35$ to $1.36$ flips the emission direction, with $F_{\mathrm{asy}}$ switching from $0.79$ to $-0.73$. The transition point can be shifted by tuning the PhC filling factor, enabling operation over different RI ranges and a larger dynamic range.

Unlike conventional PhC RI sensors that infer $n$ from small spectral shifts at fixed momentum—requiring high spectral resolution and precise angle control—our scheme relies on the radiation asymmetry, i.e., the intensity imbalance between opposite emission directions. This permits integration over a broad spectral and momentum range, potentially improving robustness and simplifying the readout. Together, these considerations suggest that the hetero-bilayer PhC provides not only switchable highly asymmetric emission, but also a promising platform for topological photonic sensing.

\begin{figure}[htb!]
\centering
\includegraphics[width=1\linewidth]{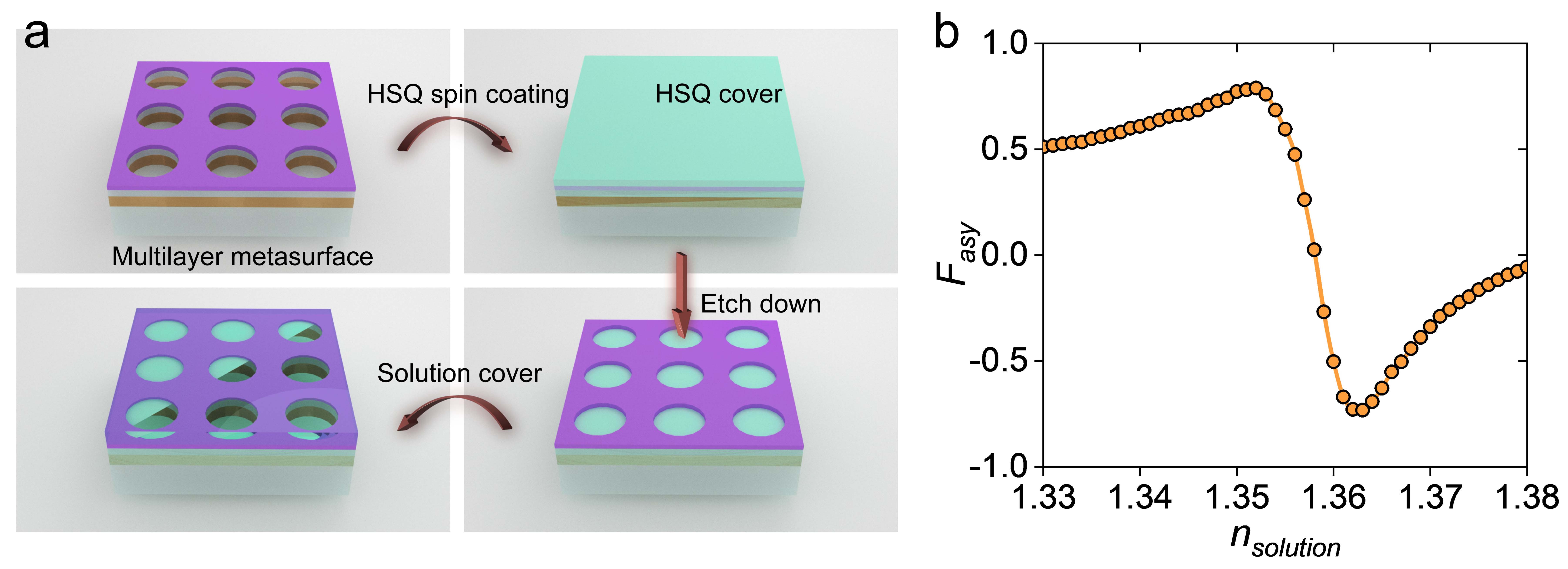}
\caption{(a) Proposed fabrication route for a dynamically switchable unidirectional-emission device. (b) Simulated RA factor versus solution refractive index for a design with $\mathrm{SiO_2}$ filling in the $\mathrm{TiO_2}$ and spacer nanoholes.}
\label{Fig.7}
\end{figure}

	
\end{document}